\documentclass[]{eptcs}

\usepackage[dvips]{graphicx}
\graphicspath{{.}}

\newcommand{\remove}[1]{}

\newcommand{\ignore}[1]{{}}

\newtheorem{THEOREM}{Theorem}

\newtheorem{COROLLARY}[THEOREM]{Corollary}

\newtheorem{LEMMA}[THEOREM]{Lemma}

\newtheorem{CLAIM}[THEOREM]{Claim}

\newtheorem{DEFINITION}{Definition}

\title{Formal Verification of an Iterative Low-Power x86 Floating-Point Multiplier with Redundant Feedback}

\author{Peter-Michael Seidel
 \institute{Advanced Micro Devices, Inc.}
 \institute{Austin, Tex., USA}
 \email{seidel@acm.org}}

\def\titlerunning{Formal Verification of an Iterative Low-power X86 Floating-point Multiplier}
\def\authorrunning{P.-M. Seidel}

\begin{document}

\maketitle 

\begin{abstract}
We present the formal verification of a low-power x86 floating-point
multiplier.  The multiplier operates iteratively and feeds back intermediate
results in redundant representation.  It supports x87 and SSE instructions in
various precisions and can block the issuing of new instructions.
The design has been optimized for low-power operation and has not been
constrained by the formal verification effort. Additional improvements for the
implementation were identified through formal verification. The formal
verification of the design also incorporates the implementation of clock-gating
and control logic.  The core of the verification effort was based on ACL2
theorem proving.  Additionally, model checking has been used to verify some
properties of the floating-point scheduler that are relevant for
the correct operation of the unit.
\end{abstract}


\section{Introduction}


Machine-assisted formal reasoning has become an integral part of the
verification strategy for many floating-point (FP) hardware designs.  At AMD,
FP hardware has already been formally verified for more than $15$
years~\cite{Russinoff05, Russinoff98mult, Russinoff95, Russinoff07cav, Russinoff09, Russinoff00mult}.

In this paper we describe the formal verification of an iterative,
low-power x86 FP multiplier that has been designed from scratch for
a new processor core. We also reiterated the verification effort for 
a second processor core that reused two instances of the design in slight 
variations with little verification overhead.
A detailed description of the initial design is presented in
\cite{Tan09}.  The unit incorporates several new features and
optimizations that cause a large number of corner cases in the design. Many of
these corner cases would have been very difficult to cover based on
only random or directed simulation.  The goal of the complete coverage
of all corner cases of the design, together with the novelty and complexity of
the design, were the initial motivation for the formal verification of this
implementation. The reuse of the verification effort for a second design 
generation has shown to be an additional benefit for this verification approach.

The core of our formal verification effort is based on ACL2 theorem
proving~\cite{ACL2}.  The unit functionality has been specified behaviorally at
RTL level in a designer-readable Verilog description~\cite{Russinoff05}. For
this specification we have also established equivalence to specifications from
previous implementations of the same FP instructions.  We built on an existing
Verilog-to-ACL2 translation tool to make the design and the specification
accessible in ACL2. We verify the design rigorously with all bit-level details
and features of the RTL that is used for production.  Additional refinements
for the implementation of certain macros are locally shown to be logically
equivalent at the implementation stage.

The FP multiplier supports x87 and SSE multiplication, reciprocal
approximation, division, and square root instructions.  Operands and results can
have one of seven different formats (two SSE, three x87, and two internal), and the
three x87 precisions can be different in the operands and in the result.  All
four rounding modes of the IEEE FP standard 754~\cite{IEEEstd754} are supported
in compliance with previous x86 implementations.  The division and square root
instructions are implemented based on a multiplicative algorithm with dedicated
use of the multiplier hardware.  Additional features for buffering,
complementing, and normalizing intermediate results, for multiply-add and
multiply-subtract operation of the multiplication array, and for the IEEE
rounding of an approximate significand quotient or square root are included in
the multiplier hardware in support of the division and square root operations.
We do not discuss these additional features in this paper, and solely focus
on the FP multiplication operation of the unit in this presentation.

Depending on the instruction and precision, FP multiplications in this unit
have a latency between two and five clock cycles. In the significand path, the
generation of the Booth-recoded partial products and their compression to a
carry-save representation of the product takes between one and three iterations
of the same stage using a rectangular $76 \times (27 + 2)$-bit multiplier
array.  A result of this stage is fed back to the previous stage in redundant
carry-save representation and added as two additional addends into the
rectangular multiplier array. For the Booth recoding of the partial products,
the result of the adder tree contains a carry bit that represents the combined
effect of the sign corrections for all partial products. The carry-save
feedback of the multiplier array may or may not contain contributions to this
carry bit.  The tracking and compensation of carries in the feedback path needs
very careful consideration of all possible scenarios. This complicates writing
cycle-based invariants for the partial-product accumulation stage.  During the
iterative partial-product accumulation, the representation of the intermediate
result is already shortened towards the target precision by iteratively
calculating a sticky and a carry bit to summarize the effect of the lower tail
of the representation on the rounding computation.  From the representation of
the partially compressed exact sum, the rounded results are then determined in
either one or two rounding stages. These rounding stages are implemented by
operation and precision, and the implementation is organized in a way very
close to having a separate rounding implementation per instruction type. The
subtle differences in the rounding algorithms for the different instruction
types significantly reduces the reuse of verification efforts between different
rounding versions.

The fact that the calculation of a multiplication can block the partial-product
accumulation stage for up to three cycles restricts the pipelining of the unit
and requires the unit to be state-controlled. New instructions may be
scheduled only when the accumulation stage is available. This condition needs to be
ensured by the FP scheduler that is implemented as a separate entity external
to the FP multiplier.  We address the correctness of the scheduling of
FP multiplication instructions to the unit with respect to the unit's
constraints as a separate verification effort.  We have addressed this separate
verification task with a different verification approach based on a commercial
model-checking tool.  We have used the definition of standard Verilog signals
to build assume-guarantee relationships between assertions and assumptions in
these two verification efforts.

The iterative nature of the implementation requires a cycle-based
setup for the control bits of the stages. In addition, operands also need to be held
appropriately to be available when they are needed (e.g., during the iterations
of the partial-product accumulation stage). The design is controlling the
clocking logic of the corresponding flops for this purpose. In the view of our
RTL translation tool, this introduces additional clocks to the design.  The
implementation of clock-gating creates a similar situation.  We deal
with both cases by removing the additional clocks and translating them to the
original clock in a pre-processing step of the translation tool.  The clock
gating logic is often dependent on reset and requires inductive
proofs. Because some of the clock-gating is implemented hierarchically,
the corresponding inductive proofs also had to be conducted hierarchically.

The design and the formal verification of this unit largely did not occur
concurrently. The design was almost completed by the time the formal
verification effort was started.  The advantage of this setting was that the
design was very stable and the verification effort had to go through very
few design modifications. One of them was the addition of a part of the clock
gating logic. But this also means the design effort was not constrained by the
formal verification effort, and the design had not been structured to simplify
modular and hierarchical specification and the application of formal
verification. This created more effort for the verification.

The formal specification and verification of the unit led to a deeper
understanding of the properties of the implementation and its operation. This
understanding allowed us to propose several improvements to the unit that were
inspired by the formal verification of some of the properties of the unit.  At
least four of these optimizations have been realized in the current
implementation.  These improvements allowed removal of some logic from the
design, reducing some critical delays in the implementation.

In Section~\ref{sec_flow} we describe our formal verification flow.
In Section~\ref{sec_spec} we describe the formal specification of the unit.
In Section~\ref{sec_design} we describe some details of the
FP multiplication implementation and 
highlight some verification challenges.
In Section~\ref{sec_results} we summarize our results before we conclude
in Section~\ref{sec_concl}.

\section{Verification Flow}   \label{sec_flow}

\begin{figure}[!t]
\centering
\includegraphics[width=2.5in]{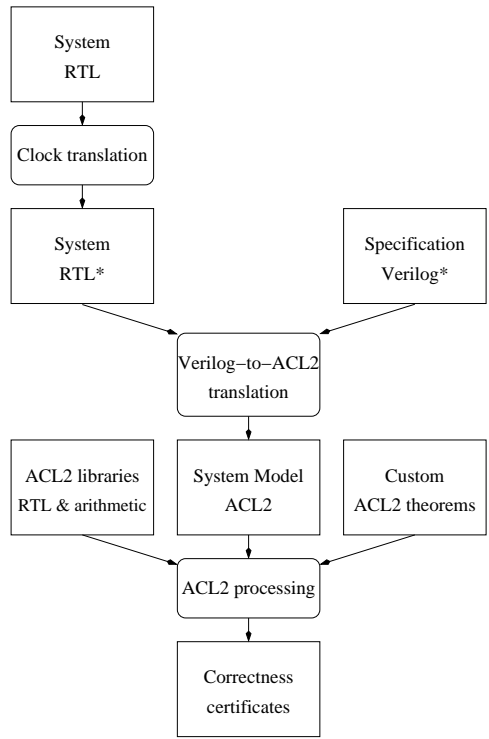}
\caption{ACL2 Verification Flow}
\label{fig_toolflow}
\end{figure}


The core of our verification is based on ACL2 theorem proving~\cite{ACL2}.  Our
ACL2 based verification flow is illustrated in Figure~\ref{fig_toolflow}.  To
verify the RTL implementation, we need to translate both the multiplier RTL and
the multiplier specification from Verilog to ACL2 logic. We use the
formalization from~\cite{Russinoff05} for the translation and build on an
existing translation tool. Just a few small extensions were necessary for the
translation tool to handle a few new features of the Verilog dialect
used in this project.  We made a few more changes to the translation
process to deal with module dependencies more efficiently and to handle the 
translation of a large number of RTL modules faster.

As a result of the translation, Verilog wires and registers are translated to 
function definitions in ACL2 that maintain the signal name as the function name
and that depend on a cycle parameter {\tt n}.  Verilog bit vectors are translated
to integer-valued ACL2 functions, and the translation process also generates
for each signal an ACL2-verified property on its bit width as expressed by its
value range.  The formalization is supported by the ACL2 RTL
library~\cite{Russinoff00mult} that has been developed during previous ACL2
verification projects at AMD and that is part of the public ACL2 distribution.
Functions to extract, concatenate, or manipulate bits and bit vectors,
and functions for logical operations, as well as a large set of verified properties
and lemmas, are provided in this library in more than $600$ function definitions
and theorems.  

In the translation, Verilog assignments are translated using the
logic definitions of the RTL library that correspond to the logic from the
Verilog assignment from the signal definition in Verilog.  The translation
process works in several stages. An early, more direct translation is
simplified in a later processing step and the equivalence of the two
translations is proven in ACL2~\cite{Kaufmann03}.  ACL2 functions for wire
definitions relate to the signals on which they depend in the same cycle. ACL2
functions for register definitions for cycle {\tt n} relate to signals from the
previous cycle {\tt n-1} or the current cycle {\tt n}.

The translation of a wire
\begin{center}
{\tt  assign  imm1  =  in1 | (in2 \& in3);}
\end{center}
results in the following ACL2 function definition for {\tt imm1}:
\begin{verbatim}
                       (DEFUN imm1 (n)
                              (LIOR (in1 n) 
                                    (LAND (in2 n) (in3 n) 
                                          1) 
                                    1))
\end{verbatim}
The translation of a register assignment
\begin{verbatim}
                       always @ (posedge clk)
                         out1  <=  in2
\end{verbatim}
results in the following ACL2 function definition for {\tt out1}:
\begin{verbatim}
                       (DEFUN out1 (n)
                              (if (zp n)
                                  (reset `out1 1)
                                  (in2 (- n 1))
                                  ))
\end{verbatim}
In this translation process, all signals are related to the common ACL2 clock
{\tt n}.  For the treatment of gated and modified clocks, we consider a
pre-processing step that translates all Verilog clocks to one common Verilog
clock, so the translation would only contain one common ACL2 clock
parameter {\tt n}.  For a gated clock {\tt gclk = gcond \& clk} and its application
to the assignment of {\tt gout1}:
\begin{verbatim}
                       always @ (posedge gclk)
                         gout1  <=  in1;
\end{verbatim}
our clock translation results in the following conditional statement
using the common clock:
\begin{verbatim}
                       always @ (posedge clk)
                         if (gcond)
                            gout1 <= in1;
\end{verbatim}
The translation of our multiplier RTL takes about $70$ minutes. This time only
needs to be spent once after each modification of the design or the
specification file to create the corresponding certified ACL2 model.  To work
on theorems using the ACL2 model, the certified model file can be loaded into
ACL2 in a matter of seconds.

\section{Formal Specification}   \label{sec_spec}

The functionality of the unit we target is the computation of FP
multiplications.  The unit supports a variety of different multiplication
options and a variety of other operations like divisions and square roots.  The
input of an opcode and an active {\tt enable} bit indicates to the unit to start
calculating on the operation represented by the opcode.

The different operations supported by the unit have different latencies. FP
multiplications take two clock cycles for packed and scalar SSE SP operation,
four clock cycles for SSE DP operation and five clock cycles for all other FP multiplications.

We would like to state as the functional specification that, if we observe the
opcode of a five-cycle multiplication in cycle {\tt n} with an active {\tt enable}
signal, the results of this instruction would be available after cycle {\tt n+4}.
\begin{verbatim}
                  (Opcode(n) == `FMUL5) & enable(n)  
                              ==> 
                  FPM.out(n+4) == FPM5spec.out(FPM.in(n))
\end{verbatim}
This is our ultimate verification goal for the five-cycle FP multiplications, and
this is what we ultimately show.  But this property is not a property of
just the FP multiplier. The unit relies on the external FP scheduler to
meet some requirements and to drive appropriate control signals to the unit.
The FP scheduler needs to avoid scheduling new instructions when the unit is
busy in iterations in its first stage and cannot accept the issue of new
instructions, or when the schedule of the new instruction would lead to
contention at the result bus of the unit, because two instructions in the unit
would finish their calculations in the same cycle. We deal with the constraints
for the FP scheduler in a separate verification effort outside of ACL2 and
reformulate the unit functionality to remove the scheduling constraints.  For
this purpose, we express the unit functionality from the observations at its
outputs and consider how these outputs have been computed.  For the output
observation in cycle {\tt n}, it could be either that two cycles ago, a two-cycle
multiplication had been started, or four cycles ago, a four-cycle multiplication
had been started, or five cycles ago a five-cycle multiplication had been started.
The following statement describes these conditions: 
\begin{verbatim}
 if ((Opcode(n-1)==`FMUL2) & enable(n-1)) 
      FPM.out(n) == FPM2spec.out(FPM.in(n-1))
 else if ((Opcode(n-3)==`FMUL4) & enable(n-3) & ~enable(n-2)) 
      FPM.out(n) == FPM4spec.out(FPM.in(n-3))
 else if ((Opcode(n-4) == `FMUL5) & enable(n-4) & ~enable(n-3) & ~enable(n-2))  
      FPM.out(n) == FPM4spec.out(FPM.in(n-4))
\end{verbatim}
This statement implicitly defines a priority for two-cycle FP multiplications
over four- and five-cycle multiplications and for four-cycle multiplications over
five-cycle multiplications, and removes any additional scheduling constraints at
the same time.  This specification statement can be shown for the FP multiplier
unit. Together with the external properties derived for the scheduler, it can
then result in the ultimate verification target from further above.

In the code statements, we used some notation that is supposed to
improve readability and represent our high-level view on the verification task.
The equations combined Verilog-like syntax with the cycle parameter {\tt n} from
the ACL2 translations.  The specification we work with in our design is
written in Verilog syntax readable to designers.  The Verilog
specification has a few extensions allowed to define rational valued registers 
and to use certified functions for the specification of IEEE-specific rounding
definitions~\cite{Russinoff00mult}.  These extensions are used to simplify and
improve specification of IEEE FP functionality.

To deal with delays in our Verilog specification, we define delayed registers 
for all inputs to the unit. If we relate {\tt enable} to the value of {\tt enable(n)} 
in cycle {\tt n} and assign
\begin{center}
\begin{verbatim}
                       always @ (posedge CLK) begin
                         enable_D1 <= enable;
                         enable_D2 <= enable_D1;
                         enable_D3 <= enable_D2;
                         enable_D4 <= enable_D3;
                       end
\end{verbatim}
\end{center}
we get the corresponding values for {\tt enable(n-1)} to {\tt enable(n-4)} in
{\tt enable\_D1} to {\tt enable\_D4}.

The arithmetic definition of the result values of the specification involves
several steps:
\begin{enumerate}
\item Extract the operand bits from register or memory format of the input operands
to the sign, exponent, and significand fields and indications of special values.
\item Define the rational values of the operands. 
\item Define the exact, unrounded (rational) operation result.
\item Define the IEEE operation result of the appropriate target precision 
      and rounding mode using parameterized IEEE rounding functions.
\item Check for value ranges and exception conditions.
\item Calculate the register format representation of the result.
\item Select the correct case for the specification result.
\end{enumerate}
An example of the definitions of rational FP operand values, of an unrounded, 
and of an IEEE-rounded result according to Steps $2$ to $4$ is given in the following:
\begin{center}
\begin{verbatim}
                 always @* begin
                   ValA = (-1) ** SignA * MantA
                          * 2 ** (ExpA - (2 ** 17 - 1) - 67);
                   ValB = (-1) ** SignB * MantB 
                          * 2 ** (ExpB - (2 ** 17 - 1) - 67);  
                   ValUnrnd  =  ValA * ValB;
                   case (RND_MODE[1:0])
                     `RN:  ValRnd  =  $Near(ValUnrnd,  Prec);     
                     `RM:  ValRnd  =  $Minf(ValUnrnd,  Prec);
                     `RP:  ValRnd  =  $Inf(ValUnrnd,   Prec);
                     `RZ:  ValRnd  =  $Trunc(ValUnrnd, Prec);
                   endcase
                 end
\end{verbatim}
\end{center}
The final specification statement in the FP multiplier specification is written
as an assert statement, which is translated into an ACL2 theorem.  The
main verification target for the ACL2 verification of FP multiplication is to verify
this main theorem. Additional theorems are targeted for exception signals and other 
output results of the unit.

The Verilog specification for the FP multiplications is largely
simplified by the behavioral features of the enhanced Verilog language
and the availability of parameterized definitions for IEEE-specific
functions (e.g., the use of the function {\tt \$Near} in extended
Verilog for the specification of IEEE rounding in rounding mode
round-to-nearest by the function {\tt near} from the ACL2 RTL
library). The writing of the specification for a larger set of x86
functions with the goal to match the functionality of previous x86
implementations of the same instructions can still be an error-prone
task.

We had the specification of previous x87 and SSE FP multiplication implementations 
available in ACL2 for the design from K7/K8~\cite{Russinoff98mult}.
Writing our specification in the same structure and using the same functions and 
definitions from the previous specification, so our Verilog specifications 
would match the previous ACL2 specifications after Verilog-to-ACL2 translation,
helped us increase confidence in the specification and in the backwards compatibility 
of our verification target.

\begin{figure}[!t]
\centering
\includegraphics[width=4.3in]{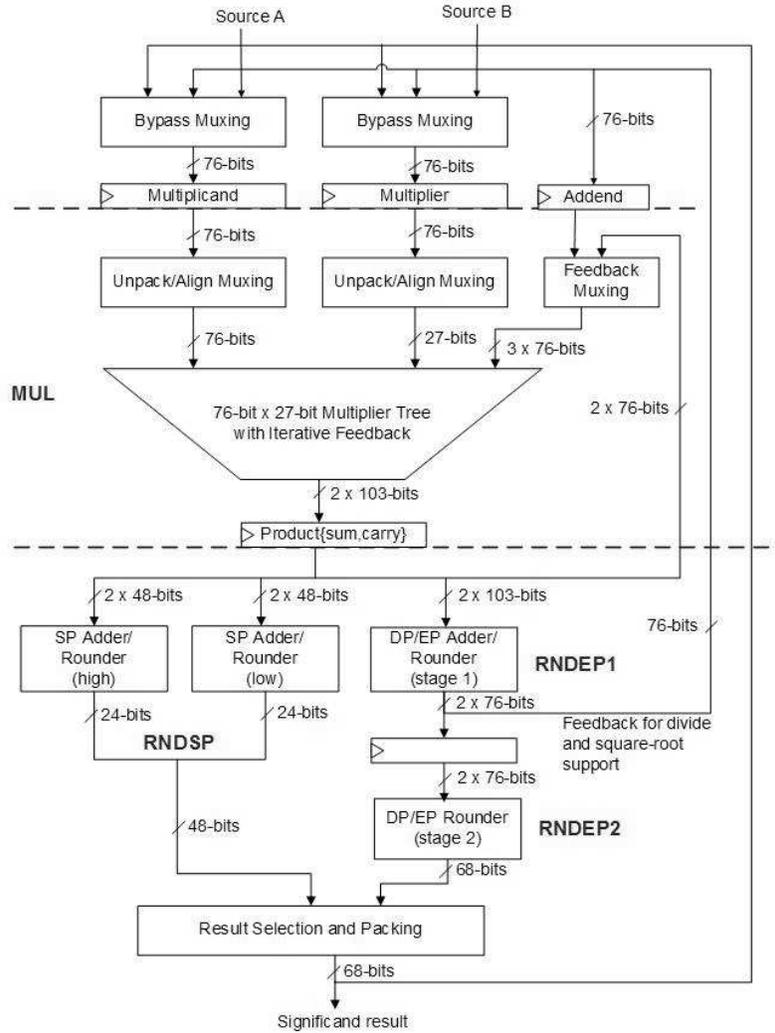}
\caption{Significand Datapath from~\cite{Tan09}}
\label{fig_sig_datapath}
\end{figure}
Our specification applied to any design evolution that we have considered.
Clock-gating did not change any of the functional behavior that was specified 
for the design. Only the theorems and proofs of properties local to the unit 
had to be adjusted to incorporate the features of the additional clock-gating logic.


\section{FP multiplier implementation} \label{sec_design}
In this section we describe some details and features of the FP multiplier
implementation. More details of the implementation can be found
in~\cite{Tan09}.  A block diagram of the FP multiplier illustrating the details
of the significand datapath is shown in Figure~\ref{fig_sig_datapath}.  To
simplify this diagram, the additional hardware for exception processing,
exponent computations, and divide/square-root support is not shown. 

The significand datapath consists of three pipeline stages. The first
pipeline stage consists of a $76 \times (27+2)$-bit multiplier that
uses modified radix-$4$ Booth recoding and a partial-product reduction
tree consisting of $4-2$ compressors. The $76 \times (27+2)$-bit
multiplier accepts a feedback product in redundant carry-save form to
facilitate iteration and a $76$-bit addend that can be added to the
product or subtracted from the product specifically to support divide
and square-root operations. The addend is needed because the
iterations for divide and square root use a restricted form of the
multiply-add operation during iterations.
The operand width of $76$ bits is required at
the micro-architectural level to support division at the internal precision of
$68$ bits that is needed for transcendental functions.  The second and third
pipeline stages consist of combined addition and rounding followed by result
selection, formatting for different precisions, and forwarding of the result to
the register file and bypass networks. 

There are two identical copies of the SP
rounding unit to support packed SP multiply operations and a single combined
DP/EP rounding unit that also handles all rounding for divide and square-root
operations. The SP rounders take one cycle and the DP/EP rounder takes two
cycles. The outputs of the two SP rounders are combined, formatted, and
multiplexed with the output from the DP/EP rounder to form the final
result. The final result is written to the register file and forwarded back to
the inputs of the FP multiplier and other FP units via the bypass networks to enhance
performance of dependent operations. With such a configuration, a scalar SP
multiplication takes one iteration, two parallel (packed) SP multiplications
take one iteration, a scalar DP multiplication takes two iterations, and a
scalar EP multiplication takes three iterations. 

The significand multiplier consists of a $76 \times (27 + 2)$-bit rectangular tree
multiplier, which performs $76 \times 76$-bit multiplications over multiple
cycles. 
This saves considerable area compared to a fully parallel $76 \times 76$-bit
multiplier, but penalizes the performance of the higher precision (DP and EP)
multiply instructions because the multiplier must stall subsequent multiply
instructions. However, the multiplier is fully pipelined for SP operations. 

The multiplier accepts a $76$-bit multiplicand input, a $76$-bit
multiplier input, and a $76$-bit addend input. These inputs are held
for the duration of the operation. The $76$-bit multiplier input is
supplied to alignment multiplexing, which outputs two $27$-bit
values. Each $27$-bit value is then recoded using a set of modified
radix-4 Booth encoders. Two separate $27$-bit multiplier values are
required to support the packed SP mode.

The outputs of the Booth encoders are used to select the multiples of
the multiplicand to form fourteen $81$-bit partial products. One of the
$27$-bit multiplier values controls the generation of the upper $38$ bits
of each partial product while the other $27$-bit multiplier value
controls the generation of the lower $38$ bits of each partial
product. In non-packed modes, the two $27$-bit multiplier values are
identical. 

\begin{figure*}[!t]
\centering
\includegraphics[width=6.5in]{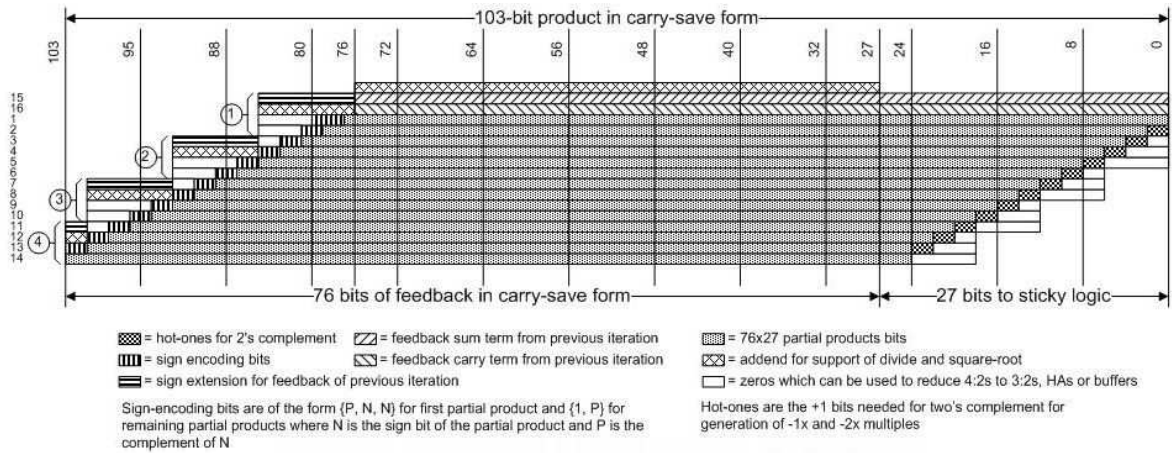}
\caption{Partial-product Layout in Modes Other than SSE SP from~\cite{Tan09}}
\label{fig_partial_products_nonSSE}
\end{figure*}

In parallel to the partial-product generation, two $76$-bit
feedback terms are combined with a $76$-bit addend using a $3-2$
carry-save adder. The $3-2$ carry-save addition is computed in parallel
with the Booth encoding and multiplexing and does not add to the
critical path. The 14 partial products plus two combined terms are
summed using a compression tree consisting of three levels of $4-2$
compressors to produce a $103$-bit product in redundant carry-save
representation. The $103$-bit carry-save product is then stored in two
$103$-bit registers.  A diagram of the partial-product array for the $76
\times 27$-bit multiplication is show in
Figure~\ref{fig_partial_products_nonSSE}.  This diagram also shows the
alignment of the two $76$-bit feedback terms and the $76$-bit addend. The
two feedback terms are needed to support iterations and are aligned to
the right. The addend is needed to support division and square root
and is aligned to the left.

To avoid unnecessary hardware, the additional terms are inserted into the unused
portions of the array wherever possible. This makes the hardware more efficient,
but also more irregular, and it adds the complexity of having to decompose and recombine
several bits and bit fields in the formulation of properties during verification. 
The ACL2 RTL and arithmetic libraries do not handle a large amount of these 
bit manipulations very efficiently.
Figure~\ref{fig_partial_products_nonSSE} also shows how the partial-product terms 
are partitioned into groups of four corresponding to the first level of
$4-2$ compressors. 
Although the multiplier is unsigned, a sign extension term is required to
accommodate the sign embedded in the uncompressed feedback terms from the
previous iteration. This is an artifact of the signed nature of the Booth
encoding and the use of sign encoding of each individual partial product.  The
two feedback terms and addend are compressed using a $3-2$ carry-save-adder (CSA)
into two terms, for a total of sixteen values to be summed.

\begin{figure*}[!t]
\centering
\includegraphics[width=6.2in]{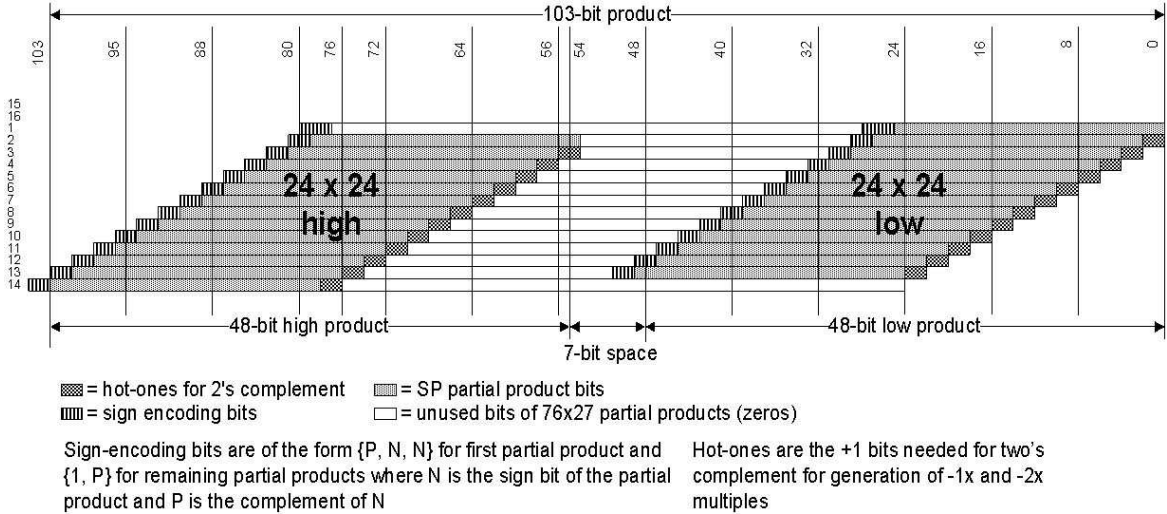}
\caption{Partial-product Layout in SSE SP Mode from~\cite{Tan09}}
\label{fig_partial_products_SSE}
\end{figure*}

To support two parallel SP multiplications, the two SP multiplications
are mapped onto the array simultaneously. The superposition of two $24
\times 24$-bit multiplier partial-product arrays onto a $76 \times
27$-bit partial-product array is shown in
Figure~\ref{fig_partial_products_SSE}. Because the lower array ends at
bit $48$, the significant bits of the upper array and lower array are
separated by seven bits. The reduction tree has three levels of $4-2$
compressors.  The width of the split between the upper and the lower
part has been justified by the designers by the number of levels of
$4-2$ compressors and based on how many bit positions a carry can
travel at most per level. Based on this reasoning, no additional
hardware had been added to kill any potential carries propagating from
the lower array into the upper array.  This kind of bit-level
justification has also been used at some other parts of the design.
It has caused the design to have some dependencies between
module-level behavioral features and bit-level implementation details
of low-level modules. These dependencies have made it more challenging
to specify the implementation in a clean, modular fashion.

To accommodate the sign encoding bits and the hot-ones, an additional
multiplexer is inserted after the Booth multiplexers and prior to the $4-2$
compressor tree. The multiplexing after the Booth
multiplexing is only required for the sign encoding bits of the lower array and
the hot-ones of the upper array, so the additional hardware required is
small. This hardware, however, is on the critical path and adds the delay of a
$2-1$ multiplexer.  

For each multiply iteration of the iterative multiplication algorithm, the
appropriate multiplier bits are selected for the high and low multiplier values
and the product is computed in redundant carry-save form. For SSE-SP multiplies
and the first iteration of all other precisions, the two feedback terms are set
to zero. For the second iteration of SSE-DP multiplies and the second and third
iterations of EP multiplies, the two feedback terms are set to the upper $76$
bits of the product from the previous iteration and are then added to the lower
$76$ bits of the current product. SP multiplies require only a single iteration,
DP multiplies require two iterations, and EP multiplies require three
iterations.
%

The rounding circuitry takes as input the product in redundant
carry-save form and rounds the result according to the given rounding
mode. The rounding circuitry contains separate rounding units for
SSE-SP high and SSE-SP low results, and a combined rounding unit that
rounds for SSE-DP, x87-EP, and divide/square-root results. Each of the
rounding units is based on a compound adder rounding scheme. The
micro-architecture requires that the FP multiplier be able to produce
the unrounded, normalized result for support of denormalized
results. This complicates the use of injection-based rounding, which
could have simplified the rounding units and their verification,
because this would be closer to an available formalization of IEEE
rounding in the ACL2 RTL library~\cite{Russinoff00mult}.  

The SSE SP rounder performs SSE single-precision rounding only. This
is a highly optimized and compact rounder compared to the DP/EP
rounder because it only has to deal with one precision. This unit has
two identical instances: one for the lower SSE-SP result and one for
the upper SSE-SP result.  In the SP rounding scheme, the upper 25 bits
are passed through one level of half-adders before applying the
compound adder. Initially, the design had implemented two levels of
half-adders for this compression. The formal analysis revealed that
the upper bound for the bits that are involved in the calculation of
the LSB of the rounded significant were smaller by one than assumed by
the designers. This allowed the removal one of the half-adder lines
from the design without adverse effects. This is one of the
optimizations that we referred to as inspired by the formal
verification effort.

The combined DP/EP rounder performs rounding for SSE-DP, x87-SP, x87-DP,
x87-EP, IP68 (for transcendental functions), and divide and square-root
operations. Due to the large number of different precisions that must be
supported, the DP/EP rounder is split over two cycles. The combined DP/EP
rounder is based on a compound adder rounding scheme. It has some similarities
with the SP rounding scheme, except it is necessary to perform a right
shift to pre-align the rounding point to the same significance prior to the
compound addition and to perform a left-shift to post-align the MSB to the same
significance after the compound addition. This is the overhead for having to support
multiple rounding points in the same datapath. 

The second difference is that the carry tree and sticky logic need to
include the carry-out and sticky from previous iterations.

The third difference is that for each target precision there is a pair
of 2-1 multiplexers that are used to insert the two rounded LSBs into
the correct positions within the final rounded significand.

The fourth difference is that for DP/EP operation, double significand
overflows can occur during rounding. The DP/EP needs to be able to
detect them, while the SP rounder can simply neglect any carries
beyond position $102$. For the DP/EP rounder, it is important to be
careful to avoid any additional carry that could be contained in the
$103$-bit carry-save representation that is fed into the rounder. The
non-existence of such carry is also a property that makes assumptions
among several module boundaries and is only justified by bit-level
details of the $4-2$ compressor implementations.  The DP/EP rounder
also provides a bypass path for divide and square root to allow the
compound adder to be reused for other additions, such as computing the
intermediate quotient $+/-1$ ULP, instead of adding dedicated
hardware.


To conclude the description of the design, we would like to point out a few selected challenges
from this verification effort. 
The implementation of the unit was new and the design had largely been completed
by the time the formal verification effort started. 
While the stability of the design was an advantage for verification, the fact that it was hard 
to justify any changes to the design to simplify the specification and verification 
effort created some challenges.

The designers were very helpful in explaining features of the design, but their 
knowledge of details and signal correlations to specify cycle- and bit-accurate 
constraints for the operation of some sub-modules was limited. Some of these constraints 
had to be determined experimentally in some iterations. 

A particular cause for complexity in the specification of some modules and
the proof of the corresponding properties was the high degree of optimization
in the implementation. The optimizations made several high-level properties dependent on
bit-level details of the design. 
Particular examples are the
carry correction logic of the redundant feedback in the adder tree iterations,
a subtle difference in the calculation of the significand overflow detection for rounding
versus the selection of the corresponding exponent adjustment, the iterative sticky 
and carry computation with logic spread over different modules,
the double significand overflow detection, and the hierarchy of clock-gating logic that is
dependent on reset.

While these features help improve the performance and lower the power of the design, 
they also complicate modular specification and formal verification with our ACL2 
theorem proving-based approach.

The design from this presentation has been reused for a second processor core
in two instances with slight modifications. The main changes in one of the instances
were related to variations in the implementation of clock-gating and the change of latencies and types
of other instructions that could be handled concurrently by the unit. The main changes
in the other instances involved the removal of RTL logic for a more efficient implementation
of a subset of the original unit's functionality. The adjustment of the verification effort
to the modified unit instances required significantly less effort than the original 
verification. But all modifications were made more complex by the properties that were
not modularized and spread over module hierarchies and boundaries. The concurrent instruction 
constraints and the clock-gating conditions could have been handled more efficiently in the 
updated instances if their properties had been better encapsulated and kept local in the 
original theorem formulations.

\section{Verification Results}  \label{sec_results}

As the main result, the ACL2 verification effort has verified the main theorem
from the Verilog specification of the unit and shown that the functionality of
the implementation meets the FP multiplication specification from K7/K8.  In
the verification of this unit, we have made use of the ACL2 RTL and arithmetic
libraries, but we also had to interactively develop and prove $8,500$ new custom
theorems and function definitions in about $250,000$ lines of LISP in $86$ files.

The time to translate the design RTL and Verilog specification to ACL2 is about
$70$ minutes; the time to certify the new theorems is about $11$ hours
on a single machine and about six hours when using multiple machines.

The design had undergone a few modifications during the verification
effort. Most of them were small and local, so it took only several days to adjust
the theorems and proof hints to the design changes. One larger design change
was the addition of an additional level of clock-gating to some parts of the
design.  This change required a significant modification in the assumptions and
invariants of several sub-units, and adjusting the theorems and proof hints to
work with the changes took a few months.

A significant number of theorems from the ACL2 verification effort
could be shown by generating some of them in a more automated way,
especially for parts of the control logic and bit-level features of
the $4-2$ compressors.  A main area for this are the properties that
are implied by control bits of the unit that have fixed values for a
specific cycle and a specific operation mode.  The propagation of
these constants through the logic and the simplification of
expressions and theorems based on these constants were needed in
multiple parts of the verification effort. In the past, we have looked into
generating properties of such propagations automatically for the case
that the values are constant during all cycles of the operation, 
but we have not adjusted this approach for the case that
independent propagations and properties, that are not generally valid, 
are to be explored for individual cycles of the operation. 
The previous effort had used byproducts of the Verilog-to-ACL2 translation process to generate
properties in ACL2 theorems that were then proven automatically.  One
way to extend this approach for cycle dependent properties could be based
on unrolling the logic for the latencies of the operations, so that
the properties would become cycle independent. Another approach could
be based on the use of more automatic features of the ACL2 theorem
prover like generating computed hints based on the propagated control
values.
The serious exploitation of these strategies did not fit into the
schedule of this verification project, but will be considered for
future ACL2-based verification at AMD.

To resolve the control and scheduling constraints for the FP multiplier, we have
used a commercial model-checking tool by Jasper Design Automation.  The main effort in this part
of the project was the reduction of the logic in the cone of influence of the
assertions. We interacted with our RTL designers for feedback on the interface constraints and
dependencies in the FP scheduler. For the remaining set of interface signals, we applied 
exhaustive exploration to discover the dependency of the assertions on the interface signals.
This helped us first to increase the depth of the search of the model checker, and finally 
to complete the proof of the assertions.  

It is hard to specify any absolute time requirement
for this part of the project because the interaction with the model checker was
run as a side project to the ACL2 verification for a larger part of the project
duration.


\section{Conclusions} \label{sec_concl}

We have discussed the formal verification of a state-of-the-art
low-power x86 FP multiplier implementation.  The unit has been
specified behaviorally in Verilog to match the functionality of
previous x86 multiplication implementations. The multiplication
implementation has been rigorously verified with all logic-level
design details including clock-gating, unit control, and the
scheduling of concurrent instructions. In this respect we have
advanced the breadth, rigor, and complexity of the formal verification
for our design and its environment compared to previous FP
multiplication verification efforts to the extent that their efforts
are revealed in the literature (e.g., ~\cite{Aagaard95,
Berg01formalverification, Kaivola02, ReeberSawada06,
Slobodova11, Slobodova04}).  The design has also incorporated several new design
features like the signed redundant iterative additive multiplier
feedback that, to the best of our knowledge, have not previously been
implemented or formally verified in production-level RTL for a
commercial FP unit implementation.

The deeper understanding of the design that was gained from the
verification process during specifying and proving design properties
has shown to be very beneficial for the current design in identifying
several improvements for the unit. We have also found that the
verification effort could be modified to variations of the design. In
two new instances of the unit that included small variations of the
design, the formal verification effort could be reused with reasonable
overhead for adjustments.  Better understanding of the challenges of
the verification process for this unit and the two modified instances
will also help make future design and verification iterations more
efficient and help identify areas of the verification process to be
targeted for improved automation.


\bibliographystyle{eptcs}

\bibliography{FVmult}


\end{document}